\documentclass[12pt,aj_pt4]{aastex}

\begin{document}
\title{ On the C/O Enrichment of Novae Ejecta }

\author{R.\ Rosner\altaffilmark{1,2},
A.\ Alexakis\altaffilmark{2},
Y.-N. Young\altaffilmark{3},
J.W.\ Truran\altaffilmark{1},
and W.\ Hillebrandt\altaffilmark{4}}

\altaffiltext{1}{Dept.\ of Astronomy \& Astrophysics, The University
of Chicago,
Chicago, IL 60637}
\altaffiltext{2}{Dept.\ of Physics, The University of Chicago, Chicago, IL
60637}
\altaffiltext{3}{Dept.\ of Applied Mathematics, Northwestern University,
Evanston, IL\ \ 60608}
\altaffiltext{4}{Max--Planck--Institut f\"ur Astrophysik,
Garching bei M\"unchen, Germany}

\begin{abstract}
Using the results of recent work in shear instabilities in stratified
fluids, we show that the resonant interaction between large-scale flows in the
accreted H/He envelope of white dwarf stars and interfacial gravity waves can
mix with the star's envelope with the white dwarf's surface material, leading
to the enhancement of the envelope's C/O abundance to levels required by extant
models for nova outbursts.
\end{abstract}


The substantial enrichment of CNO nuclei in the ejecta from novae (cf.\ Truran
1985; Gehrz et al.\ 1998 and references therein) has been a puzzle for over two
decades. Early theoretical models of nova outbursts (e.g., Starrfield, Truran
\& Sparks 1978; Fujimoto 1982) clearly showed that nuclear processing during
hydrogen burning during the nova flash could not account for the observed CNO
abundances, which can reach 30\% by mass. These early studies already
recognized that the solution to the puzzle must involve ``dredgeup" of C/O from
the white dwarf before or during the nova outburst. This mixing was required
both to meet the constraints on CNO abundances in the ejecta and to power the
nova itself, since the energy production rate per unit mass depends directly on
the metallicity (e.g., Wallace \& Woosley 1981); thus, Starrfield et al.\
(1978) and Fujimoto (1982) showed explicitly with 1-D models that runaway in a
pure H/He envelope did not release enough energy in order to eject enough
matter with sufficient velocity to match observations.

Concerns that mixing may occur at the interface between the accreting
matter and
the underlying star (and already accreted material) followed closely upon the
recognition that such mixing was essential in order to understand the elemental
composition of the nova ejecta (Starrfield et al.\ 1972). At that time, there
was already some interest in understanding mixing at the interface between a
stellar surface and an accreting flow. For example, Kippenhahn \& Thomas (1978)
examined shear flow instability in the stratified boundary layer between a
white dwarf and the infalling accretion flow associated with an accretion disk;
and established the linear stability properties (based on using the Richardson
number\footnote{The Richardson number, $Ri = \alpha g A l_o / U_o^2$, is a
measure of the competition between the stabilizing effect of buoyancy and the
destabilizing effect of the shear flow. Here $\alpha$ is the coefficient of
volume expansion, $g$ is the local gravitational acceleration,
$l_o$ is a local characteristic length scale, and $U_o$ is the shear flow
amplitude; $A$ is the Atwood number across the material interface separating
the white dwarf surface and the H/He envelope [$\equiv (\rho_{\rm star} -
\rho_{\rm envelope}) / (\rho_{\rm star} + \rho_{\rm envelope})$].} as the
control parameter).

The shear instability considered by Kippenhahn \& Thomas (also Sung 1974)
has been extensively revisited (viz., MacDonald 1983). Kippenhahn
\& Thomas conjectured that this instability saturates at the marginal state for
stability, and therefore weak mixing; MacDonald, upon revisiting this problem,
showed that the shear instability would lead to rapid dispersal of the accreted
matter over the entire white dwarf surface (as opposed to the relatively narrow
accretion belt which emerged from Kippenhahn \& Thomas' analysis), but also
suggested that the radial mixing time was long (set by the thermal timescale of
the envelope). These arguments lead to rather minimal mixing, and for these
reasons shear mixing has not been regarded as a likely candidate for the
required mixing process.\footnote{However, very recently, Br\"uggen \&
Hillebrandt (2001a,b) have begun to re-examine the nonlinear aspects of this
problem computationally, in an attempt to place the earlier analytical
calculations on a firmer quantitative footing.} Indeed, until the late 1990's
the absence of a plausible mixing mechanism was considered to be a major
stumbling block for understanding novae. In the mid-90's, several authors
conjectured that the convection which was known to initiate some $\sim$ 1,000
years before runaway might be associated with convective undershoot and
convective penetration, processes which might lead to mixing of the stellar C/O
into the envelope (Shankar, Arnett, \& Fryxell 1992; Shankar \& Arnett 1994),
but quantitative calculations were not done until the mid and late 1990's
(Glasner \& Livne 1995; Glasner et al.\ 1997; Kercek, Hillebrandt, \& Truran
1998a). These more recent calculations investigated the possibility that
convective undershoot just before, and possibly during, nova runaway might lead
to the required mixing. However, Kercek, Hillebrandt, \& Truran (1998b, 1999)
have shown convincingly (both by comparing two and three-dimensional
simulations, and by conducting resolution studies in which the extent of mixing
was measured as a function of grid resolution) that convective undershoot was
not likely to work as an effective mixing process. In particular, the
resolution studies showed {\sl less} mixing as grid resolution was increased.
This can be readily understood if the boundary layer between the stellar
surface and the accreted (convecting) envelope is laminar: in that case, since
the dominant viscosity in these simulations is numerical, increased resolution
leads to a thinner boundary (or mixing) layer, and whence to {\sl less}
mixing as the grid resolution is increased. Thus, it would appear that we are
once again lacking an effective mixing process.

For this reason, we have recently reexamined the physics of shear flow
instabilities (Alexakis, Young, \& Rosner 2001; Young et al.\ 2001). The
question to answer was whether previous astrophysical studies of this subject
had in fact fully explored this mixing process. As we show below, the past work
in fact missed an important aspect of shear mixing in stratified media. In the
following, we will extract the critical aspects of our earlier results that
apply to the problem at hand.

One very important aspect of the shear mixing problem is that the instability
that leads to turbulent mixing between fluid layers depends only on certain
essential features of the shear flow. For present purposes, it suffices to
consider prototypical velocity profiles of the form
$U(z) = U_o + U_1 \ln (z / \sigma +1)$ for $z \ge 0$
($U(z)=0$ below the interface), or
$U(z) = U_o + U_1 \tanh (z / \sigma )$ for $z \ge 0$
($U(z)=0$ below the interface),
where $U_o$ is the velocity jump (if any) at the envelope/stellar surface
interface, $z$ is the vertical coordinate (with $z=0$ marking the
initial envelope-star interface), and $\sigma$ is the characteristic
scale length of the shear flow in the envelope.\footnote{The logarithmic
velocity profile is commonly observed in the boundary layer of winds blowing
over the surface of extensive bodies of water [cf.\ Miles 1957]; the tanh
profile has the advantage of bounded shear velocity far from the shear
interface.} It is then well-known that if $U_o=0$, the Kelvin-Helmholtz
instability is entirely absent for a velocity profile with either of
the two functional forms given above. Since (in the presence of viscosity) the
relative velocity  between the two layers of fluid at the interface must be
zero (i.e., an attached flow boundary condition) and since $\sigma$ is
proportional to the viscous boundary layer thickness, Kelvin-Helmholtz
instability is unlikely to be important in contributing to the mixing on
relevant spatial and time scales for the above wind profile, which appears to
be a  reasonable approximation to the actual boundary layer flow (cf., Alexakis
et al.\ 2001).

Now, the same argument applies to the generation of terrestrial surface water
waves by winds; and as originally pointed out by Miles (1957; see also Phillips
1957), winds are nevertheless able to amplify such surface waves to finite
amplitude. Hence, there must be some other instability present. One such
instability (critical-layer instability) was identified and studied extensively
in the linear regime by Miles (1957), Howard (1961), Lighthill (1962), and
others. This instability originates from the continuum of unstable modes formed
when surface gravity waves travel at the same velocity as the wind at some
height above the interface. In order to treat this instability, it is essential
to formulate the shear instability problem more generally; this is in part the
motivation for considering velocity profiles of the forms given above; velocity
profiles of this type in stratified atmospheres have been explored extensively
by the geophysical fluid dynamics community. These previous
geophysically-motivated studies were largely confined to the parameter regime
characteristic of the water/air interface; but recently Alexakis, Young, \&
Rosner (2001) have fully explored the control parameter space governing these
instabilities: these nondimensional parameters are the Atwood number
$A$ and the
Richardson number $Ri$ (see Alexakis et al.\ 2001); in physical terms, the key
parameters are the gravitational acceleration $g$, the shear scale length
$\sigma$, the density ratio $\rho_{\rm ‡star}/\rho_{\rm envelope}$, and the
shear amplitude $U$. In this Letter, we now apply the results presented by
Alexakis et al.\ to the astrophysical context by developing a new model for the
interface mixing.

First, consider the underlying physics: This is most readily done in
the context
of a particularly simple model for the interface, in which the shear flow has
step discontinuity across the density interface between the C/O white dwarf
surface and the bottom of the H/He envelope. Start with the simplest case, in
which we ignore stratification. The classical Kelvin-Helmholtz instability is
then based on the observation that a spatial interface perturbation can be
destabilized because the flow must speed up over the ``hills" of the
perturbation and slow over the ``valleys"; Bernoulli's law then tells us that a
low-pressure region develops over the ``hills", and a high-pressure region over
the ``valleys", thus pulling up the ``hills" and pushing down the ``valleys",
leading to a linear instability whose growth rate $\gamma ~ \sim kU$, where $k$
is the wavenumber of the interface perturbation and $U$ is the shear amplitude.

If the shear flow interface is not a step, but has finite thickness
(viz., given
by $\sigma$, as above), and if stratification is allowed, then it
is well-known that the dispersion relation is no longer linear (Chandrasekhar
1962, \S 102), and that both low and high wavenumber cutoffs appear, with
$\gamma > 0$ only for $\zeta_{\rm min} < k \sigma < \zeta_{\rm max}$ (where the
values of $\zeta_{\rm min}$ and $\zeta_{\rm max}$ depend on the specifics of
the velocity profile (cf.\ Figures 119 and 120 in Chandrasekhar 1962), and
maximum growth at, for example, $(k \sigma )^2 \sim 0.5$ for the
tanh$(z/\sigma)$ shear profile. As a result, instability can only occur in a
finite region of the wavenumber-Richardson number plane; for the tanh velocity
profile, the stability boundary is defined by the curves  $J = 0$ and $J = (k
\sigma)^2 (1 - (k \sigma)^2 )$, with instability only in the domain bounded by
these two curves. In order to apply this to the nova case, we simplify the
actual case by assuming an exponentially-stratified background
atmosphere of the
form $\rho (z) \sim \rho_o \exp({-\beta z})$, with a horizontal shear layer of
form $U = U_o \tanh (z/ \sigma )$ located at the white dwarf surface; it is
readily seen that in the event that this surface shear flow is driven by
thermal convection in the overlying envelope, then the unstable modes will lie
in a wavelength band defined by
$
6 \times 10^4 ~{\rm cm}< \lambda_{\rm unstable} < 2 \times 10^5 ~(T/10^8 ~{\rm
K})^{1/2}~ (U_o/10^5~{\rm cm~s^{-1}})
$ cm,
where we have assumed a white dwarf of radius $\sim 10^{-2} R_{\rm Sun}$, a
shear layer thickness $\sigma \sim 10^4$ cm, gravitational acceleration $g_{\rm
wd} \sim 2.7 \times 10^8 ~{\rm cm~s^{-2}}$, and density scale height
$\beta^{-1} \sim 3 \times 10^{8} (T/10^8~K) $ cm; inclusion of the density jump
at the (C,O/H,He) interface would lower the wavelength of unstable modes yet
further. The upper bound on this mixing scale is of order the grid
resolution in
the currently highest-resolution calculations (viz., Glasner, Livne \& Truran
1997; Kercek, Hillebrandt \& Truran 1999), consistent with the observation by
these authors that little shear mixing occurred in their computations. Since
there is an upper bound on the shear flow length scale $\sigma$ in order for
Kelvin-Helmholtz instability to occur at all\footnote{This bound is computed
from the stability criterion $J < 1/4$ for the tanh velocity profile; thus
$\sigma < 1.67 \times 10^4~(T/10^8~{\rm K})^{1/2} (U_o/10^5~{\rm cm~s}^{-1})$
cm in order for Kelvin-Helmholtz instability to occur at all.}, and since the
mixing scale is at most of order 10 times the shear scale, this suggests that
Kelvin-Helmholtz instability will not be an effective CNO mixing process under
any circumstances.

In contrast, consider the interaction of the same wind with the normal
modes supported by the interface between the stellar surface and the accreted
envelope. These normal modes are akin to ``deep water waves" seen at the
surface of terrestrial oceans; and are known to grow in amplitude as a result
of the resonant interaction between these waves and the wind. More
specifically, at any given wavenumber $k$, linear theory provides the wave's
phase velocity
$
v_{\rm phase} \equiv \omega / k \sim (A/k)^{1/2} (g + \Sigma k^2 / \rho^{\rm
water})
\sim (Ag/k)^{1/2}
$,
where $\Sigma$ is the surface tension; in the case of the gaseous media
characterizing stars, the surface tension term is of course absent. For any
given wind profile, $U(z)$, where $z$ is the vertical coordinate, one can then
satisfy a resonance between the wind and a surface mode such that
$
U(z) = v_{\rm phase} \sim (Ag/k)^{1/2}
$;
that is, a wave with wavenumber $k \sim Ag/U(z)^2$ will be driven resonantly
unstable. (For typical values of $a$, $g$, and $U$ characteristic of a white
dwarf surface, one finds that the wavelength of unstable modes lies
in the range
0.01 -- 1 km.)\ \ The key issue is then how to determine the mixing layer width
once these unstable modes cease their growth and finally saturate: naively, one
might expect the saturation process to simply limit the mode amplitude, and
thereby determine the width of the mixing layer. In the case of interfacial
gravity modes, however, saturation is well-known to occur via wavebreaking (see
Chen, Kharif, Zaleski, \& Li 1999 and references therein); it is the resulting
spray that then determines (from a statistical point of view) the
effective mean
width of the mixing layer -- this width can be substantially larger than the
mode amplitude at saturation, as is well-known in the case of wind-driven spray
from breaking ocean waves. In any case, let us assume for the moment that we
have determined this layer width, which we shall denote as $\lambda$. Finally,
we note that while one would need in general to take into account
stratification effects (viz., molecular weight gradients) on either side of the
density jump, such effects are to lowest order unimportant here because the
mixing layer defined by wave breaking is likely to be much narrower than the
local gravitational scale height.

We are now ready to describe our simplified model: consider first the amount
of carbon and oxygen in the breaking wave mixing layer, which we write in the
form
$
M_{\rm C+O}^{\rm mixing~layer} \sim \alpha \rho_0^{\rm envelope}
\lambda \Lambda
\xi
$,
where $\alpha$ is the coefficient for the C+O mass fraction in the
mixing layer,
$\lambda$ is the mixing layer width (both $\alpha$ and $\lambda$ are to be
determined from simulations; cf.\ Young et al.\ 2001), $\rho^{\rm envelope}_0$
is the density of the envelope at its base (i.e, in the breaking wave mixing
layer), $\Lambda$ is the characteristic length scale of the large-scale
circulation (which can be identified with the outer scale of convection in the
envelope); and $\xi$ is a length scale transverse to the wind direction (this
dimension will drop out of our formulation). Note that the remaining parameters
appearing in this relation can be obtained from extant (1-D) nova models. Now,
as mentioned earlier, the amount of C+O needed to be mixed into the envelope is
roughly 1/3 by mass of the ejecta mass $M^{\rm envelope}_{\rm total}$, or
$
M_{\rm C+O}^{\rm envelope}\sim (1/3) M^{\rm envelope}_{\rm total} \sim
(1/3) \cdot \left( (2/3) \Lambda \rho^{\rm envelope}_0 \cdot \Lambda \xi
\right)
$.
The ``sweepout time", i.e., the time scale on which the boundary mixing layer
is swept out by a penetrating convective roll, is just
$
\tau_{\rm sweep} \sim \Lambda / U
$,
so that the time needed to mix the necessary amount of carbon and oxygen into
the envelope is just
$
\tau_{\rm mixing} \sim M_{\rm C+O}^{\rm envelope}/(M_{\rm
C+O}^{\rm mixing~layer}/\tau_{\rm sweep})
$,
or
\begin{eqnarray*}
\tau_{\rm mixing} & \sim & (2/9) \Lambda^2 / \alpha \lambda U \\
& \sim & 5 \alpha^{-1} (\Lambda/10^8 {\rm cm})^2 (\lambda / 10^2 {\rm cm})^{-1}
\cdot \\
& & ~~\cdot (U/10^5 {\rm cm~s^{-1}})^{-1} ~{\rm years},
\end{eqnarray*}
with $\alpha \sim 0.3 - 1$. Thus, it is evident that the evolution time scale
for the envelope prior to nova runaway (which is roughly of the order of the
time between onset of envelope convection and runaway, or $\sim 10^3$
years) is much longer than the mixing timescale. This confirms that
resonantly-driven mixing at the star-envelope boundary can be an efficient
mixing process during the pre-nova star evolution; the clear next step is to
verify these results via simulations of weakly compressible fluids subject to
these mixing instabilities. We also note that this mixing time scale is much
longer than the dynamical time characteristic of the nova runaway itself. For
this reason, the amount of additional C+O material mixed in during the outburst
itself can be regarded as a small perturbation. One remaining significant issue
relates to the possible effects of magnetic fields on the C+O mixing process;
that is, one might be concerned that turbulent mixing may be suppressed if
local magnetic fields in the envelope become large as convection sets on $\sim$
1,000 years before runaway. We are not currently in a position to resolve this
possible problem, but only note that because the conservative mixing time scale
$\tau_{\rm mixing} << 1,000$ yrs, substantial mixing suppression by magnetic
fields could be accommodated within this model without vitiating the
main point,
namely that resonant instability of the C+O/envelope boundary can lead to
effective mixing across that boundary. This is a critical point for any nova
model because novae have been observed for white dwarfs with relatively strong
magnetic fields (e.g., V1500 Cygni 1975; Stockman, Schmidt \& Lamb 1988).
However, in the absence of a detailed calculation, this point remains to be
addressed by theory.

To conclude, by using the results of linear stability theory, as well as
extrapolating from existing numerical simulations of nova outbursts, we have
estimated the mixing zone parameters, and have shown that pre-nova erosion of
the wave-breaking mixing layer by slow convection could mix sufficient C/O into
the accreted H/He envelope to satisfy observations. We have constructed a
simple mixing length subgrid prescription to describe this mixing process, and
have shown that this subgrid model only needs to be used for the pre-nova
phase. Further mixing during the outburst is no longer required. Because the
C/O abundance in the envelope builds up gradually during the pre-nova slow
convective phase, we expect the nova envelope mass attained before outburst may
be substantially larger than in standard models assuming a ``pre-seeded"
envelope.

\acknowledgements
This work has been supported by the DOE-funded ASCI/Alliances Center for
Astrophysical Thermonuclear Flashes at the University of Chicago.


\end{document}